\documentclass[aps,showpacs,preprint,preprintnumber,nofootinbib,amsmath,amssymb,ascmac,12pt]{revtex4}
\usepackage{bm}
\usepackage[dvipdfm]{graphicx}
\begin{document}
\title{\vspace*{0.5cm}
A simple diagnosis of non-smoothness of black hole horizon
\\
{\it Curvature singularity at horizons in extremal Kaluza-Klein black holes
}}
\author{
${}^{1}$Masashi Kimura\footnote{E-mail:m.kimura@damtp.cam.ac.uk},
${}^{2}$Hideki Ishihara\footnote{E-mail:ishihara@sci.osaka-cu.ac.jp},
${}^{2}$Ken Matsuno\footnote{E-mail:matsuno@sci.osaka-cu.ac.jp}
and
${}^{3,4}$Takahiro Tanaka\footnote{E-mail:t.tanaka@tap.scphys.kyoto-u.ac.jp},
}

\affiliation{\vspace*{0.5cm}
${}^{1}$DAMTP, University of Cambridge, Centre for Mathematical Sciences,
Wilberforce Road, Cambridge CB3 0WA, UK
\\
${}^{2}$Department of Mathematics and Physics, Osaka City University, Sumiyoshi, Osaka 558-8585, Japan
\\
${}^{3}$Department of Physics, Kyoto University, Kyoto 606-8502, Japan
\\
${}^{4}$
Yukawa Institute for Theoretical Physics, Kyoto University, 
Kyoto 606-8502, Japan
}
\begin{abstract}
We propose a simple method to prove non-smoothness of a black hole horizon.
The existence of a $C^1$ extension across the horizon
implies
that there is no $C^{N + 2}$ extension across the horizon if some components of 
$N$-th covariant derivative of Riemann tensor diverge at the horizon in the coordinates of the $C^1$ extension.
In particular, the divergence of a component of the Riemann tensor 
at the horizon directly indicates the presence of a curvature singularity.
By using this method, we can confirm the existence of a curvature singularity for several cases 
where the scalar invariants constructed from the Riemann tensor,
{\it e.g.}, the Ricci scalar and the Kretschmann invariant, 
take finite values at the horizon.
As a concrete example of the  application,
we show that the Kaluza-Klein black holes constructed by Myers have a curvature singularity at the horizon 
if the spacetime dimension is higher than five.
\end{abstract}

\preprint{OCU-PHYS 406}
\preprint{AP-GR 112}
\preprint{KUNS-2507}
\preprint{YITP-14-60}


\date{\today}
\maketitle

\section{Introduction}
\label{introduction}
To test whether our world is the higher-dimensional spacetime,
we need to identify phenomena which clearly indicate the existence of extra dimensions.
Recently, the study of higher dimensional black holes has attracted much attention
under the expectation that they may 
have characteristic features of extra dimensions.
For example, in the higher dimensional scenarios
based on the TeV gravity 
mini black holes might be produced 
in a linear collider~\cite{Banks:1999gd, Dimopoulos:2001hw, Giddings:2001bu, Ida:2002ez, Ida:2005ax, Ida:2006tf} 
or in cosmic ray events~\cite{Argyres:1998qn, Feng:2001ib, Anchordoqui:2001cg} unlike the case of four dimensional gravity.

Since the sizes of extra dimensions should be compact from a realistic point of view,
we should study higher dimensional spacetime with compactified extra dimensions, 
{\it i.e.}, Kaluza-Klein (KK) spacetime.
In this paper, we focus on higher dimensional black holes in KK spacetime~(KK black holes).
As a first step, it would be important to investigate exact solutions of KK black holes
to understand their qualitative feature.
In the five dimensional case, recent studies showed that there exist 
a variety of KK black holes called 
squashed KK black holes~\cite{Dobiasch:1981vh, Gibbons:1985ac, Gauntlett:2002nw, Gaiotto:2005gf, 
Ishihara:2005dp, Wang:2006nw, Yazadjiev:2006iv, Nakagawa:2008rm, Tomizawa:2008hw, Matsuno:2008fn, 
Tomizawa:2008rh, Stelea:2008tt, Tomizawa:2008qr, Gal'tsov:2008sh, Bena:2009ev, Tomizawa:2010xq, 
Mizoguchi:2011zj, Chen:2010ih, Stelea:2011fj, Nedkova:2011hx, Tatsuoka:2011tx, Nedkova:2011aa, 
Mizoguchi:2012vg}.
However, in general, to construct an exact solution of KK black hole is difficult 
because of the less symmetry except for special cases.
In fact, if $D \ge 6$ and the number of the non-compact dimensions is four,
the KK black holes constructed from multi black holes solutions~\cite{Myers:1986rx} 
are the only family of exact solution of KK black holes with spherical topology.

Though one might think that we cannot construct an exact solution of multi black holes 
since the gravitational force is only attractive, 
it is possible if each black hole has the same mass and 
charge\footnote{In this case, the black hole horizon becomes extremal.},
where the gravitational attractive force is balanced with the Coulomb 
force.
Such exact solutions are known as 
Majumdar-Papapetrou solutions~\cite{Majumdar:1947eu, Papaetrou:1947ib, Hartle:1972ya},
and then higher-dimensional generalization was considered by Myers~\cite{Myers:1986rx}.
In Ref.~\cite{Myers:1986rx}, 
Myers constructed KK black holes by placing an infinite number of 
black holes
in a lattice configuration, which is equivalent to placing a single black hole 
with an appropriate periodic identification of space.

In Ref.~\cite{Candlish:2007fh} Candlish and Reall showed that higher dimensional 
multi black holes have a non-smooth event horizon\footnote{
The non-smoothness of the horizon was firstly investigated in~\cite{Gibbons:1994vm, Welch:1995dh},
and re-investigated in detail~\cite{Candlish:2007fh}.
In~\cite{Candlish:2009vy} and~\cite{Kimura:2008cq},
the case of rotating black hole and non trivial topology~\cite{Ishihara:2006iv, Ishihara:2006pb}
were studied, respectively.
Recently, the case of multi-center coplanar black hole and
membrane horizons were studied in~\cite{Gowdigere:2014aca}.
}
unlike the case of four dimensional multi black holes, which have an analytic horizon~\cite{Hartle:1972ya}.
In $D=5$ we can find a $C^2$ (but not $C^3$) extension 
across the horizon.
By contrast, in $D \ge 6$  the metric is not $C^2$ but $C^1$ at the horizon since 
some components of 
the second derivatives of the metric always diverge at the horizon for any extension across the horizon.
This means the existence of a curvature singularity at the horizon in $D \ge 6$.

The result in Ref.~\cite{Candlish:2007fh} seems to indicate that 
KK black holes constructed by Myers~\cite{Myers:1986rx} also have a curvature singularity at the horizon 
because they are constructed from higher dimensional multi black holes.
However it is non trivial
whether this expectation is correct or not 
since we consider an infinite number of black holes in the case of KK black holes,
which 
might be qualitatively different from the case of a finite number of black holes.
In fact, as shown in Ref.~\cite{Candlish:2007fh},
five dimensional KK black holes with a $S^1$ compactified extra dimension 
have an analytic event horizon\footnote{
In $D\ge 6$, KK black holes with a $S^1$ compactified extra dimension 
do not have $C^2$ horizon~\cite{Candlish:2007fh}.}
in contrast to the case of a finite number of black holes, whose horizon is not $C^3$.
Extrapolating the results in $D=4$ and $D=5$, 
we can expect that 
the horizon might be analytic
if the number of the non-compact dimensions is four.
Therefore,
we would like to investigate whether or not
KK black holes with $T^{D-4}$ compactified extra dimensions can have a smooth horizon in $D \ge 6$.
However, since the methods used in previous works are restricted to the axi-symmetric case~\cite{Candlish:2007fh} 
or coplanar case~\cite{Gowdigere:2014aca},
we need to develop a new tool to investigate the  smoothness of KK black holes with less symmetry.

In this paper, as an approach to this issue,  
we propose a simple method to prove non-smoothness of a black hole horizon which applies to
less-symmetric cases.
Our claim is
that there is no $C^{N + 2}$ extension across the horizon if some components of 
$N$-th covariant derivative of Riemann tensor diverge at the horizon 
in the coordinates of a $C^1$ extension across the horizon.
Furthermore,
the divergence of a component of the Riemann tensor means that
a curvature singularity appears on the horizon.
Using this method, we can identify a curvature singularity even when
the scalar invariants constructed from the Riemann tensor,
{\it e.g.}, the Ricci scalar and the Kretschmann invariant, 
are finite at the horizon.
As an application of this method,
we show that the Kaluza-Klein black holes constructed 
by Myers have a curvature singularity at the horizon 
in $D \ge 6$.

This paper is organized as follows. 
In Sec.~II, we develop a method to prove non-smoothness of horizon.
In Sec.~III, we 
apply our method to a spherically symmetric toy model and 
show our method works well.
In Sec.~IV, we discuss the case of multi black holes and show 
our method can reproduce the result of previous works.
We discuss the case of Kaluza-Klein black holes and show that there exists
curvature singularity on the horizon if $D \ge 6$ in Sec.~V.
Sec.~VI is devoted to the summary and discussion.
We use the units in which $c = G = 1$.

\section{method to prove non-smoothness of horizon}
\label{formalism}

Black hole solutions are usually constructed 
in a single coordinate system
which does not cover the event horizon.
If we want to discuss the global structure of such solutions,
we need to find an extension across  coordinate boundaries,
such as an event horizon.
If the spacetime admits an analytic extension,
we can find a unique and natural extension of the original spacetime.
However, as shown in the previous works~\cite{Gibbons:1994vm, Welch:1995dh, Candlish:2007fh, 
Candlish:2009vy, Gowdigere:2014aca, Chrusciel:1992tj},
some black hole spacetimes are not smooth at the horizon,
but in general it is not easy to prove the non-smoothness of the horizon.
In this section, we develop a method 
which can be applied to prove non-smoothness of black hole horizon.

\begin{figure}[htbp]
\begin{center}
\includegraphics[width=0.4\linewidth,clip]{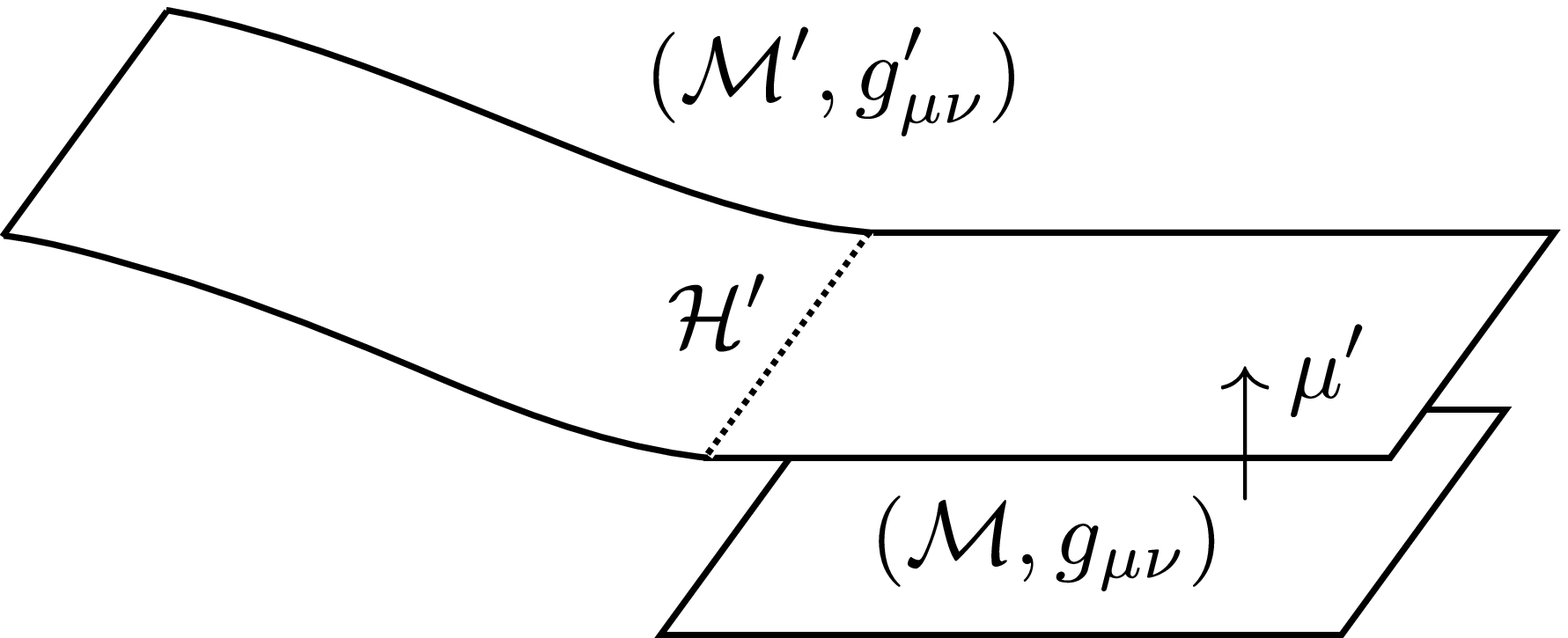}
\end{center}
\caption{
$C^1$ extension $({\cal M^\prime}, g_{\mu \nu}^\prime)$ of $({\cal M},g_{\mu \nu})$.
}
\label{fig1:extension}
\end{figure}

Let $({\cal M},g_{\mu \nu})$ be a $D$-dimensional $C^\infty$ manifold with a $C^\infty$ metric tensor with Lorentzian signature
and $({\cal M^\prime}, g_{\mu \nu}^\prime)$ be a $D$-dimensional 
extension of $({\cal M},g_{\mu \nu})$
with an isometric imbedding $\mu^{\prime} : {\cal M} \to {\cal M^\prime}$.
Here ${\cal M^\prime}$ is a
$C^2$ manifold~\footnote{
In this paper, we assume all manifolds are at least $C^2$ 
so that we can consider the second derivative of a geodesic curve w.r.t. affine parameter.}
and the metric $g_{\mu \nu}^\prime$ is $C^1$
on this manifold.\footnote{
The submanifold $\mu^{\prime}({\cal M})$ can be considered as $C^\infty$ manifold 
with $C^\infty$ metric
because of the existence of isometric imbedding $\mu^{\prime} : {\cal M} \to {\cal M^\prime}$.
}
We assume that 
the boundary of $\mu^{\prime}({\cal M})$
contains a smooth hypersurface ${\cal H^\prime}$ in ${\cal M^\prime}$.\footnote{
A hypersurface ${\cal H^\prime}$ is smooth if 
it is described by an equation $\phi = 0$ 
where $\phi$ is a $C^1$ function on ${\cal M^\prime}$ and
it satisfies $d \phi \neq 0$ on ${\cal H^{\prime}}$.
For example, in the case of four-dimensional Schwarzschild spacetime $ds^2 = -(1-2M/r)dt^2 + (1-2M/r)^{-1}dr^2 + r^2 d\Omega^2$, 
we can consider an extension as 
$ds^2 = -(1-2M/r)du^2 + 2dudr + r^2 d\Omega^2$.
In this case, ${\cal H^\prime}$ corresponds to the horizon $r = 2M$.
If we choose $\phi$ as $\phi = r- 2M$, $\phi$ satisfies that $\phi = 0$ and $d\phi \neq 0$ at the horizon.
}
(see Fig.\ref{fig1:extension})

At first, we define {\it an extension of $({\cal M}, g_{\mu \nu})$
across the ``same'' boundary} for later convenience.
Let $({\cal \bar{M}}, \bar{g}_{\mu \nu})$
be an another $D$-dimensional $C^1$ extension
 of $({\cal M},g_{\mu \nu})$ with an
isometric imbedding $\bar{\mu} : {\cal M} \to {\cal \bar{M}}$.
We assume the boundary of $\bar{\mu}({\cal M})$ also contains a smooth hypersurface $\bar{\cal H}$
in $\bar{\cal M}$.
Let $\gamma(\lambda) : (\lambda_i, \lambda_f) \to {\cal M}$ be an incomplete geodesic with 
affine parameter $\lambda$
in ${\cal M}$ such that $\gamma(\lambda \to \lambda_f) \notin {\cal M}$ and
$\mu^\prime(\gamma(\lambda))$ reaches a point in ${\cal H}^\prime$, 
{\it i.e.}, there exists a limit point 
$p^\prime = \mu^\prime(\gamma(\lambda \to \lambda_f)) \in {\cal H}^\prime$.
If $\bar{\mu}(\gamma(\lambda))$ also reaches a point in ${\cal \bar{H}}$, {\it i.e.},
if there exists a limit point 
$\bar{p} = \bar{\mu}(\gamma(\lambda \to \lambda_f)) \in \bar{\cal H}$,
for all such $\gamma(\lambda)$,
 $({\cal M^{\prime}}, g_{\mu \nu}^{\prime})$ and
 $({\cal \bar{M}}, \bar{g}_{\mu \nu})$
are called the extensions  of $({\cal M}, g_{\mu \nu})$
across the ``{\it same}'' boundary.
\\

In this setup we would like to prove the following theorem.
\\

{\it Theorem.1}~~{\it
Let $({\cal M},g_{\mu \nu})$ be a $D$-dimensional $C^\infty$ manifold with a $C^\infty$ metric tensor with Lorentzian signature
and $({\cal M^\prime}, g_{\mu \nu}^\prime)$ be a $D$-dimensional 
extension of $({\cal M},g_{\mu \nu})$ 
with an isometric imbedding $\mu^{\prime} : {\cal M} \to {\cal M^\prime}$.
We assume that ${\cal M^\prime}$ is a
$C^2$ manifold, the metric $g_{\mu \nu}^\prime$ is $C^1$
on this manifold
and the boundary of $\mu^{\prime}({\cal M})$
contains a smooth hypersurface ${\cal H^\prime}$ in ${\cal M^\prime}$.
If at least one component of the
Riemann tensor diverges at a point on ${\cal H}^\prime$
independently of the approaching direction from $\mu^\prime({\cal M})$
in the coordinates of the $C^1$ extension $({\cal M^\prime}, g_{\mu \nu}^\prime)$,
there is no
$C^2$ extension of $({\cal M}, g_{\mu \nu})$ across the ``same'' boundary.}
\\

We divide the proof of this theorem into two steps, lemmas.~1 and 2.
{}From these lemmas, we can immediately prove the above theorem.
Roughly speaking, the lemma.~1 implies that
if there exists a parallelly propagated~(p.p.) curvature singularity
at the boundary of $({\cal M},g_{\mu \nu})$
along a time like geodesic, then there is no $C^2$ extension in which
the geodesic is extendible across the boundary.
The lemma.~2 claims 
the existence of p.p. curvature singularity
at the boundary of $\mu^\prime({\cal M})$
 along a time like geodesic
under the same assumption as theorem.~1.
\\

{\it Lemma.1}~~{\it 
Let $({\cal M},g_{\mu \nu})$ be a $D$-dimensional $C^\infty$ manifold with a $C^\infty$ metric tensor with Lorentzian signature.
If $({\cal M},g_{\mu \nu})$ has a 
p.p.~curvature singularity
at the boundary of $({\cal M},g_{\mu \nu})$
along a time like geodesic with finite affine parameter,
{\it i.e.},
there exists an incomplete time like geodesic 
$\gamma(\lambda):(\lambda_{i}, \lambda_{f}) \to {\cal M}$
 with an affine parameter $\lambda$ in ${\cal M}$
such that $\gamma(\lambda \to \lambda_f) \notin {\cal M}$
and at least one component of the Riemann tensor measured by
a parallelly propagated frame along $\gamma(\lambda)$
diverges in the limit $\lambda \to \lambda_{f}$,
then
there is no $D$-dimensional
$C^2$ extension of $({\cal M}, g_{\mu \nu})$ 
such that
the boundary of the isometric imbedding map 
of ${\cal M}$ contains a smooth hypersurface
and 
the map of $\gamma(\lambda)$ has a limit point on the hypersurface 
in the limit $\lambda \to \lambda_{f}$.}
\\

{\it Proof.}~~
Suppose that 
there also existed a $D$-dimensional extension $({\cal M^{\prime \prime}}, g_{\mu \nu}^{\prime \prime})$
 of $({\cal M}, g_{\mu \nu})$ with
 an isometric imbedding $\mu^{\prime \prime} : {\cal M} \to {\cal M^{\prime \prime}}$
such that ${\cal M^{\prime \prime}}$ is a $C^2$ manifold and the metric 
$g_{\mu \nu}^{\prime \prime}$ is $C^2$ on ${\cal M^{\prime \prime}}$,
and 
the time like geodesic $\gamma^{\prime \prime}(\lambda) := \mu^{\prime \prime}(\gamma(\lambda))$ 
reaches a point on the smooth hypersurface ${\cal H^{\prime \prime}}$
which is contained in the boundary of $\mu^{\prime \prime}({\cal M})$.
Let $p^{\prime \prime}=\gamma^{\prime \prime}(\lambda \to \lambda_f)
 \in \partial(\mu^{\prime \prime}({\cal M}))$ be the point on ${\cal H^{\prime \prime}}$.
Using local coordinates $\{y^\mu \}$ around $p^{\prime \prime}$, 
we denote the geodesic $\gamma^{\prime \prime}(\lambda)$ in ${\cal M^{\prime \prime}}$ by $y^\mu(\lambda)$.
The geodesic equation for $y^\mu(\lambda)$ becomes
\begin{eqnarray}
\frac{d^2y^\mu}{d\lambda^2} &=& 
-\Gamma^\mu{}_{\alpha \beta} \frac{dy^\alpha}{d\lambda}\frac{dy^\beta}{d\lambda},
\label{geoeqc2extension}
\end{eqnarray}
with the Christoffel symbols $\Gamma^\mu{}_{\alpha \beta}$.
We can show that 
the tangent $dy^\mu/d\lambda$ does not diverge at the point $p^{\prime \prime}$ 
if $\gamma^{\prime \prime}(\lambda)$ is a time like geodesic
as discussed in Appendix~\ref{appendixA}.
Thus, we can uniquely extend the geodesic across ${\cal H^{\prime \prime}}$.

Let $e_{(\alpha)}^\mu$ denote the linearly independent 
parallelly propagated vectors along $\gamma^{\prime \prime}(\lambda)$, 
where the index $(\alpha)$ distinguishes different vectors, and 
we assume $e_{(0)}^\mu$ is tangent to $\gamma^{\prime \prime}(\lambda)$.
Then, the components of $e_{(\alpha)}^\mu$ measured by the coordinate basis of $\{y^\mu \}$
take finite value at $p^{\prime \prime}$ as shown in~Appendix~\ref{appendixB}.

On the other hand, we have a relation 
\begin{eqnarray}
 R_{(\alpha) (\beta) (\gamma) (\delta)} = 
e_{(\alpha)}^\mu e_{(\beta)}^\nu e_{(\gamma)}^\rho e_{(\delta)}^\sigma 
R_{\mu \nu \rho \sigma}^{(y^\mu)},
\label{tetradcomponent0}
\end{eqnarray}
where $R_{(\alpha) (\beta) (\gamma) (\delta)}$ and 
$R_{\mu \nu \rho \sigma}^{(y^\mu)}$
are
the components of the Riemann tensor in the basis of $e_{(\alpha)}^\mu$ and
in the coordinate basis of $\{y^\mu\}$, respectively.
{}From the assumption, at least one component of 
$R_{(\alpha) (\beta) (\gamma) (\delta)}$ diverges at ${\cal H^{\prime \prime}}$, while
$R_{\alpha \beta \gamma \delta}^{(y^\mu)}$ remains finite there.
Thus at least one component of $e_{(\alpha)}^\mu$ must diverge at ${\cal H^{\prime \prime}}$.
However, this contradicts the result 
established in the preceding paragraph
that $e_{(\alpha)}^\mu$ are finite. 
\hfill$\Box$
\\

{\it Lemma.2}~~{\it
Let $({\cal M},g_{\mu \nu})$ be a $D$-dimensional $C^\infty$ manifold with a $C^\infty$ metric tensor with Lorentzian signature
and $({\cal M^\prime}, g_{\mu \nu}^\prime)$ be a $D$-dimensional 
extension of $({\cal M},g_{\mu \nu})$ 
with an isometric imbedding $\mu^{\prime} : {\cal M} \to {\cal M^\prime}$.
We assume that ${\cal M^\prime}$ is a
$C^2$ manifold
and the metric $g_{\mu \nu}^\prime$ is $C^1$
on this manifold
and the boundary of $\mu^{\prime}({\cal M})$
contains a smooth hypersurface ${\cal H^\prime}$ in ${\cal M^\prime}$.
If at least one component of the
Riemann tensor diverges at a point on ${\cal H}^\prime$
independently of the approaching direction from $\mu^\prime({\cal M})$
in the coordinates of the $C^1$ extension $({\cal M^\prime}, g_{\mu \nu}^\prime)$,
there exists a p.p.~curvature singularity at a point $p^\prime$ on ${\cal H^\prime}$
along a time like geodesic reaching from $\mu^\prime({\cal M})$ with finite affine parameter.}
\\

{\it Proof.}~~
Since the Christoffel symbols are finite, we can always move to coordinates in which
the metric is apparently locally flat at the point $p^\prime$ on ${\cal H^\prime}$.
Then, we can
prepare linearly independent orthonormal basis vectors
$v^{\mu}_{(\alpha)}$ at $p^\prime$ such that
satisfy
\begin{eqnarray}
v^\mu_{(\alpha)} v_{\mu(\beta)} = \eta_{(\alpha)(\beta)} = {\rm diag}[-1,1,\cdots,1],
\end{eqnarray}
and we assume 
that $v^\mu_{(0)}$ 
is not parallel to ${\cal H^\prime}$.
Solving the geodesic equation from ${\cal H^\prime}$ with the initial velocity $v^\mu_{(0)}$,
and considering parallel transport of $v^\mu_{(\alpha)}$ along this geodesic, we obtain
an orthonormal frame spanned by $e_{(\alpha)}^\mu$ along the geodesic.
Without loss of generality, we can assume
that $v^\mu_{(0)}$ is a past-directed time like vector such that 
the geodesic stays in $\mu^\prime({\cal M})$
as long as a sufficiently short geodesic is concerned.\footnote{
If the geodesic does not stay in $\mu^\prime({\cal M})$
even for a sufficiently short geodesic,
we only have to change $v^\mu_{(0)}$ into $-v^\mu_{(0)}$ and
 interchange the past and the future in the following discussion.}
(See Appendices~\ref{appendixC} and \ref{appendixB} for the existence of the solution of geodesic equation and
the orthonormal vectors $e_{(\alpha)}^\mu$, respectively)

A component of the Riemann tensor in
the vielbein frame spanned by 
$e_{(\alpha)}^\mu$ becomes
\begin{eqnarray}
 R_{(\alpha) (\beta) (\gamma) (\delta)} = 
e_{(\alpha)}^\mu e_{(\beta)}^\nu e_{(\gamma)}^\rho e_{(\delta)}^\sigma 
R_{\mu \nu \rho \sigma}.
\label{tetradcomponent}
\end{eqnarray}
Since the vielbein frame is linearly independent, 
$e_{(\alpha)}^\mu$ has its inverse $e^{-1}{}_\mu^{(\alpha)}$
defined by 
\begin{eqnarray}
\sum_\alpha e_{(\alpha)}^\mu e^{-1}{}_\nu^{(\alpha)} &=& \delta^\mu_{\nu},
\end{eqnarray}
we can rewrite Eq.~(\ref{tetradcomponent}) as
\begin{eqnarray}
\sum_{\alpha,\beta,\gamma,\delta}
e^{-1}{}_\mu^{(\alpha)} e^{-1}{}_\nu^{(\beta)} e^{-1}{}_\rho^{(\gamma)} e^{-1}{}_\sigma^{(\delta)}
 R_{(\alpha) (\beta) (\gamma) (\delta)} = 
R_{\mu \nu \rho \sigma}.
\label{tetradcomponent2}
\end{eqnarray}
{}From the assumption, at least one component of 
$R_{\mu \nu \rho \sigma}$ diverges at $p^\prime$.
As shown in Appendix~\ref{appendixB}, $e_{(\alpha)}^\mu$ and $e^{-1}{}_\mu^{(\alpha)}$
take finite values at $p^\prime$, 
and hence
$R_{(\alpha) (\beta) (\gamma) (\delta)}$ must diverge at $p^\prime$
along the geodesic.
\hfill$\Box$
\\

We can easily generalize the above theorem
to be able to 
prove no existence of $C^{N+2}$ extension across the same boundary.
\\

{\it Theorem.2}~~{\it
Let $({\cal M},g_{\mu \nu})$ be a $D$-dimensional $C^\infty$ manifold with a $C^\infty$ metric tensor with Lorentzian signature
and $({\cal M^\prime}, g_{\mu \nu}^\prime)$ be a $D$-dimensional 
extension of $({\cal M},g_{\mu \nu})$ 
with an isometric imbedding $\mu^{\prime} : {\cal M} \to {\cal M^\prime}$.
We assume that ${\cal M^\prime}$ is a
$C^2$ manifold, the metric $g_{\mu \nu}^\prime$ is $C^1$
on this manifold
and the boundary of $\mu^{\prime}({\cal M})$
contains a smooth hypersurface ${\cal H^\prime}$ in ${\cal M^\prime}$.
If at least one component of the
$N$-th covariant derivative of the
Riemann tensor diverges at a point on ${\cal H}^\prime$
independently of the approaching direction from $\mu^\prime({\cal M})$
in the coordinates of the $C^1$ extension $({\cal M^\prime}, g_{\mu \nu}^\prime)$,
there is no
$C^{N+2}$ extension of $({\cal M}, g_{\mu \nu})$ across the ``same'' boundary.}
\\

{\it Proof.}~~
Suppose that 
there also existed a $C^{N + 2}$ extension 
$({\cal M^{\prime \prime}}, g_{\mu \nu}^{\prime \prime})$
 of $({\cal M}, g_{\mu \nu})$.
First, 
similarly to the discussion in the proof of the lemma.~2, we can say that
the divergence of the $N$-th covariant derivative of the 
Riemann tensor at a point on ${\cal H}^\prime$
in the coordinates of a $C^1$ extension 
independently of the approaching direction 
implies 
that at least one component of the 
$N$-th covariant derivative of the Riemann tensor 
measured by a parallelly propagated frame along 
a time like geodesic diverges at a point on ${\cal H}^\prime$.
Next, similarly to the discussion in the proof of the lemma.~1,
we can say that
such a divergence of 
the $N$-th covariant derivative of the Riemann tensor 
measured by a parallelly propagated frame implies the divergence of the
$N$-th covariant derivative of the Riemann tensor 
in the coordinate basis of $C^1$ extension. 
However this contradicts our assumption.
\hfill$\Box$
\\

\section{Application I : spherically-symmetric toy model}
In this section, as an example, 
we consider a deformed Schwarzschild spacetime in four dimensions
\begin{eqnarray}
ds^2 &=& -f dt^2 + f^{-1}dr^2 + 
r^2 \left(1 + \frac{m^{1/2}(r-2m)^{3/2}}{r^2}\right)(d\theta^2 + \sin^2\theta d\phi^2),
\\
f &=& 1 - \frac{2m}{r},
\end{eqnarray}
and we apply our method to this spherically-symmetric toy model, 
and show that there exists a curvature singularity at the horizon $r = 2m$.

First, to obtain $C^1$ extension across the horizon, we introduce a new coordinate $u$ as
\begin{eqnarray}
dt = du - f^{-1}dr.
\end{eqnarray}
Then, the metric becomes
\begin{eqnarray}
ds^2 = -f du^2 
+2 dudr + r^2 \left(1 + \frac{m^{1/2}(r-2m)^{3/2}}{r^2}\right)(d\theta^2 + \sin^2\theta d\phi^2).
\label{toymodelefcoord}
\end{eqnarray}
We can easily check that this is a $C^1$ extension but not a $C^2$ extension across the horizon
because of the existence of the factor $(r-2m)^{3/2}$.

Though the Ricci scalar $R$ and the Kretschmann invariant 
$R^{\mu \nu \rho \sigma}R_{\mu \nu \rho \sigma}$
take finite values at the horizon, for this metric,  
we can show that a component of 
the Riemann tensor in the coordinates of $C^1$ extension~(\ref{toymodelefcoord}) behaves as
\begin{eqnarray}
R_{r\theta r\theta} \simeq -\frac{3}{8}\frac{\sqrt{m}}{\sqrt{r-2m}} \to -\infty.
\end{eqnarray}
{}From the theorem.~1,
the divergence of Riemann tensor in the coordinate of $C^1$ extension
implies that there exists no $C^2$ extension across the horizon at $r = 2m$ and 
that there always exists a curvature singularity at the horizon.

In this symmetric case, we can also show that there exists the p.p.~curvature singularity
in the usual way.
We prepare the orthogonal vielbein bases $e^{(\mu)} = e^{(\mu)}_{\alpha} dx^\alpha$ as
\begin{eqnarray}
e^{(0)} &=& -dt - \frac{\sqrt{2m r}}{r -2 m} dr,
\notag\\
e^{(1)} &=& - \frac{\sqrt{2mr}}{r}dt - \frac{r}{r - 2m} dr,
\notag\\
e^{(2)} &=& 
r \left(1 + \frac{m^{1/2}(r-2m)^{3/2}}{r^2}\right)^{1/2}d\theta,
\notag\\ 
e^{(3)} &=& 
r \left(1 + \frac{m^{1/2}(r-2m)^{3/2}}{r^2}\right)^{1/2} \sin \theta d\phi,
\end{eqnarray}
such that $e^{(\mu)}_\alpha$ satisfy $g^{\alpha \beta} e^{(\mu)}_\alpha e^{(\nu)}_\beta 
=\eta^{\mu \nu} = {\rm diag}[-1,1,1,1]$ and 
$ g^{\alpha \beta} e^{(0)}_{\alpha} \nabla_\beta e^{(0)}_{\gamma} = 0$.
Then we obtain
\begin{eqnarray}
e_{\alpha}^{(1)}
e_{\beta}^{(2)}
e_{\gamma}^{(1)}
e_{\delta}^{(2)}
R^{\alpha \beta \gamma \delta} \simeq - \frac{3}{32 m^{3/2}\sqrt{r-2m}} \to - \infty.
\end{eqnarray}
We can see that the p.p.~curvature singularity exists at the horizon $r = 2m$
and
the behavior of the Riemann tensor 
in the coordinates of the $C^1$ extension
is basically the same as the vielbein component of the Riemann tensor for a free fall observer.

\section{Application II : case of multi black holes}
In this section we apply our method to the case of multi black hole solutions
to reproduce the results of previous works~\cite{Welch:1995dh, Candlish:2007fh}.

\subsection{construction of multi black holes}
We consider $D$-dimensional Einstein-Maxwell system described by the action
\begin{eqnarray}
S = \frac{1}{16\pi G_D}\int d^Dx \sqrt{-g}(R - F_{\mu \nu} F^{\mu \nu}),
\end{eqnarray}
where 
$R$ is the Ricci scalar, $F_{\mu \nu} = \partial_\mu A_\nu -  \partial_\nu A_\mu$ is the Maxwell field strength,
and
$G_D$ is $D$ dimensional gravitational constant.
From this action we obtain the Einstein equations and the Maxwell equations as
\begin{eqnarray}
R_{\mu \nu} - \frac{1}{2}R g_{\mu \nu}
&=& 2\left(
F_{\mu \lambda}F_{\nu}{}^{\lambda}
-\frac{1}{4}g_{\mu \nu}F_{\rho \sigma}F^{\rho \sigma}
\right),
\label{einsteineq}
\\
\nabla_\nu F^{\mu \nu} &=& 0.
\label{maxwelleq}
\end{eqnarray}
In this paper, as a solution of Eqs~(\ref{einsteineq}) and (\ref{maxwelleq}),
we focus on $D$-dimensional Majumdar-Papapetrou solution~\cite{Majumdar:1947eu,Papaetrou:1947ib,Myers:1986rx},
whose metric and gauge 1-form are
given by
\begin{eqnarray}
ds^2 &=& -H^{-2} dt^2 + H^{2/(D-3)} \sum_{i,j=1}^{D-1} \delta_{ij}dx^i dx^j,
\label{MPsol}
\\
A_{\mu}dx^\mu &=& \pm \sqrt{\frac{D-2}{2(D-3)}} H^{-1} dt,
\label{MPsolgaugepotential}
\\
H &=& 1 +  \sum_{n} \frac{m_n}{|\bm{x} - \bm{a}_n|^{D-3}},
\end{eqnarray}
where $\bm{x}$ and $\bm{a}_n$ 
denote the position vector and 
the location of the horizon of the $n$-th black hole in
$D-1$ dimensional Euclid space, respectively.
At $\bm{x} = \bm{a}_n$ the lapse $g_{tt}$ vanishes.
The mass parameter of the $n$-th black hole is denoted by $m_n$.

In this section, for simplicity, we focus on the case of two black holes.
The metric becomes
\begin{eqnarray}
ds^2 &=& -H^{-2} dt^2 + H^{2/(D-3)} (dr^2 + r^2 d\theta^2 + r^2 \sin^2\theta d\Omega_{S^{D-3}}^2),
\label{twobhmetric1}
\\
H &=& 1 + \frac{m_1}{r^{D-3}}
+
\frac{m_2}{(r^2 - 2 ar \cos{\theta} + a^2)^{(D-3)/2}},
\end{eqnarray}
where $d\Omega_{S^{D-3}}^2$ is the metric of an unit $D-3$ sphere and
$a$ specifies the separation between the two black holes. 
We set the horizon of one black hole to the origin of Euclidean space.

\subsection{$C^1$ extension across the horizon}
At the horizon $r=0$, the metric component $g_{rr}$ 
in Eq.~(\ref{twobhmetric1}) diverges.
To remove this divergence, we first introduce the Eddington-Finkelstein coordinate $u$ as
\begin{eqnarray}
dt &=:& du - H^{(D-2)/(D-3)}dr + Y d\theta,
\label{EFcoordtwobh1}
\\
Y &=& - \int dr \partial_\theta H^{(D-2)/(D-3)}.
\end{eqnarray}
The last term in Eq.~(\ref{EFcoordtwobh1}) is needed to satisfy integrability condition.
Near the horizon, the metric behaves as
\begin{eqnarray}
ds^2  &\simeq &  
 2 m_1^{-(D-4)/(D-3)} r^{(D-4)} dr du 
+
m_1^{2/(D-3)} (d\theta^2 + {\sin}^2 \theta d\Omega_{S^{D-3}}^2).
\label{EFcoordtwobh2}
\end{eqnarray}
If $D > 4$, 
the metric in this coordinate degenerates at the horizon since 
$g_{ru}$ vanishes at $r = 0$.
We can remove this coordinate singularity by introducing new radial coordinate $\rho$ as
\begin{eqnarray}
\rho = r^{D-3},
\end{eqnarray}
However, since the function $H$ becomes
\begin{eqnarray}
H &=& 1 + \frac{m_1}{\rho}
+
\frac{m_2}{(\rho^{2/(D-3)} - 2 a\rho^{1/(D-3)} \cos{\theta} + a^2 )^{(D-3)/2}},
\end{eqnarray}
we find that the last term in the denominator contains fractional powers of $\rho$. Thus this extension 
is not an analytic extension across the horizon.
Even worse, since the metric behaves as
\begin{eqnarray}
ds^2 &=&
 2 m_1^{-(D-4)/(D-3)} (D-3)^{-1}
 du d\rho + 2F d\rho d\theta 
\notag\\ &&
+m_1^{2/(D-3)}  (d\theta^2 + {\sin}^2 \theta d\Omega_{S^{D-3}}^2)
+ 
{\cal O}(\rho),
\\
F &=&
(D-3)^{-1}  m_1^{-(D-4)/(D-3)} Y 
\notag\\&= &
(D-3)^{-1}(D-2)  a^{-(D-2)}  m_1^{-(D-5)/(D-3)}
 m_2
 \sin \theta \left(
\rho^{1/(D-3)} 
+
{\cal O}(\rho^{2/(D-3)})
\right),
\end{eqnarray}
the first derivative of $g_{\rho \theta}$ diverges at the horizon
because of the fractional power of $\rho$ in the form of $F$.
Then, this is only $C^0$ extension.
To obtain a $C^1$ extension, we finally introduce new coordinates 
$\bar{u}$ and $\bar{\theta}$ as
\begin{eqnarray}
d\bar{\theta} &:=& d\theta + \frac{1}{m_1^{2/(D-3)}} F d\rho
+\frac{1}{m_1^{2/(D-3)}}d\theta \int d\rho \partial_\theta F,
\label{thetabartwobh}
\\
d\bar{u} &:=& m_1^{-(D-4)/(D-3)} (D-3)^{-1}du -   \frac{1}{m_1^{2/(D-3)}} \frac{F^2}{2} d\rho
- \frac{1}{m_1^{2/(D-3)}} d\theta \int d\rho \partial_\theta \frac{F^2}{2}.
\label{ubartwobh}
\end{eqnarray}
Notice that the leading terms of the last terms in the Eqs.~(\ref{thetabartwobh}) and (\ref{ubartwobh}) 
have positive powers of $\rho$ higher than unity because their integrands 
are proportional to positive fractional powers of $\rho$ near $\rho = 0$.
Then the metric behaves as
\begin{eqnarray}
ds^2 =
 2  d\bar{u} d\rho
+ m_1^{2/(D-3)}(d\bar{\theta}^2 + {\sin}^2 \bar{\theta} d\Omega_{S^{D-3}}^2)
+{\cal O }(\rho^1).
\end{eqnarray}
We can see that
the derivatives of all the metric component w.r.t. $\rho$ take finite values at the horizon,
namely, this is a $C^1$ extension across the horizon.

\subsection{Divergence of Riemann tensors in the coordinate of $C^1$ extension}
We should comment that two coordinate bases
$(du, d\rho, d\theta)$ and $(d\bar{u}, d\rho, d\bar{\theta})$ are 
linearly related with non-degenerate finite coefficients at $\rho = 0$.
Thus, if a component of a tensor diverges at $\rho = 0$ in the coordinates $(du, d\rho, d\theta)$,
at least one component of the tensor
diverges at $\rho = 0$ also in the coordinates $(d\bar{u}, d\rho, d\bar{\theta})$. 
For this reason, 
it is sufficient to confirm the divergence of the Riemann tensor in the coordinates $(du, d\rho, d\theta)$
in order to conclude its divergence in the coordinates of the $C^1$ extension $(d\bar{u}, d\rho, d\bar{\theta})$.

In five dimensional case, we can easily verify that  all the components of the Riemann tensor 
in the coordinates $(du, d\rho, d\theta)$ take finite values at the horizon.
We can also show a component of the first covariant derivatives behaves as
\begin{eqnarray}
\nabla_{\rho} R_{\rho u \theta \rho} 
\propto \frac{1}{\rho^{1/2}}
\to \infty.
\end{eqnarray}
Thus, we conclude that there is no $C^3$ extension in $D=5$ from the theorem.~2.
As for $C^2$ extension, 
we can construct such an extension by using the Riemann normal coordinates.\footnote{
Although the metric has only $C^1$ on the horizon, 
we can still construct a Riemann normal coordinate
by using a solution of the geodesic equation in Appendix.~\ref{appendixB}.}

If the dimension is higher than five, the Riemann tensor behaves as
\begin{eqnarray}
R_{\rho \theta \rho \theta} \propto \frac{1}{\rho^{(D-5)/(D-3)}}
\to \infty.
\end{eqnarray}
Thus, we conclude that there is no $C^2$ extension across the horizon from the theorem.~1.
and 
that we cannot remove the curvature singularity on the horizon by considering 
any extension across the horizon.

\section{Application III : case of Kaluza-Klein black holes}
In the case of $D \ge 5$,
if we superpose black holes with the same mass $m$ periodically,
we can obtain 
a toy model of a Kaluza-Klein black hole with $T^N$ compactified extra 
dimensions as constructed by Myers~\cite{Myers:1986rx} by using multi 
black hole solutions Eqs.~(\ref{MPsol}) and (\ref{MPsolgaugepotential}).

In this section we apply our method to this Kaluza-Klein black hole solution.
One might think that 
the curvature singularity should exist at the horizon 
since it exists even in the two black hole case.
However, whether this expectation is correct or not is not so obvious. 
Infinite superposition of black holes may have qualitatively different feature of spacetime 
than the case of finite number superposition.

In fact, the previous works show that
a five dimensional Kaluza-Klein black hole admits analytic extension
across the horizon~\cite{Candlish:2007fh}
in contrast with the case of two black holes where the horizon is not $C^3$ but $C^2$.
The four dimensional multi black hole solutions also admit analytic extension across the horizon.
Hence, we can also expect that the Kaluza-Klein black hole admits smooth extension across the horizon
when the number of non-compact dimensions is four.
In general, such a Kaluza-Klein black hole spacetime becomes less-symmetric.
While the method used in previous works~\cite{Candlish:2007fh, Gowdigere:2014aca} can be applied 
only to axi-symmetric or plane symmetric case, 
our method is applicable to this case.

In this section, we clarify whether the Kaluza-Klein black hole 
admits smoother extension compared to the case of two black hole
when the number of non-compact dimensions is four.
\subsection{metric form of Kaluza-Klein black holes}
\begin{figure}[htbp]
\begin{center}
\includegraphics[width=0.4\linewidth,clip]{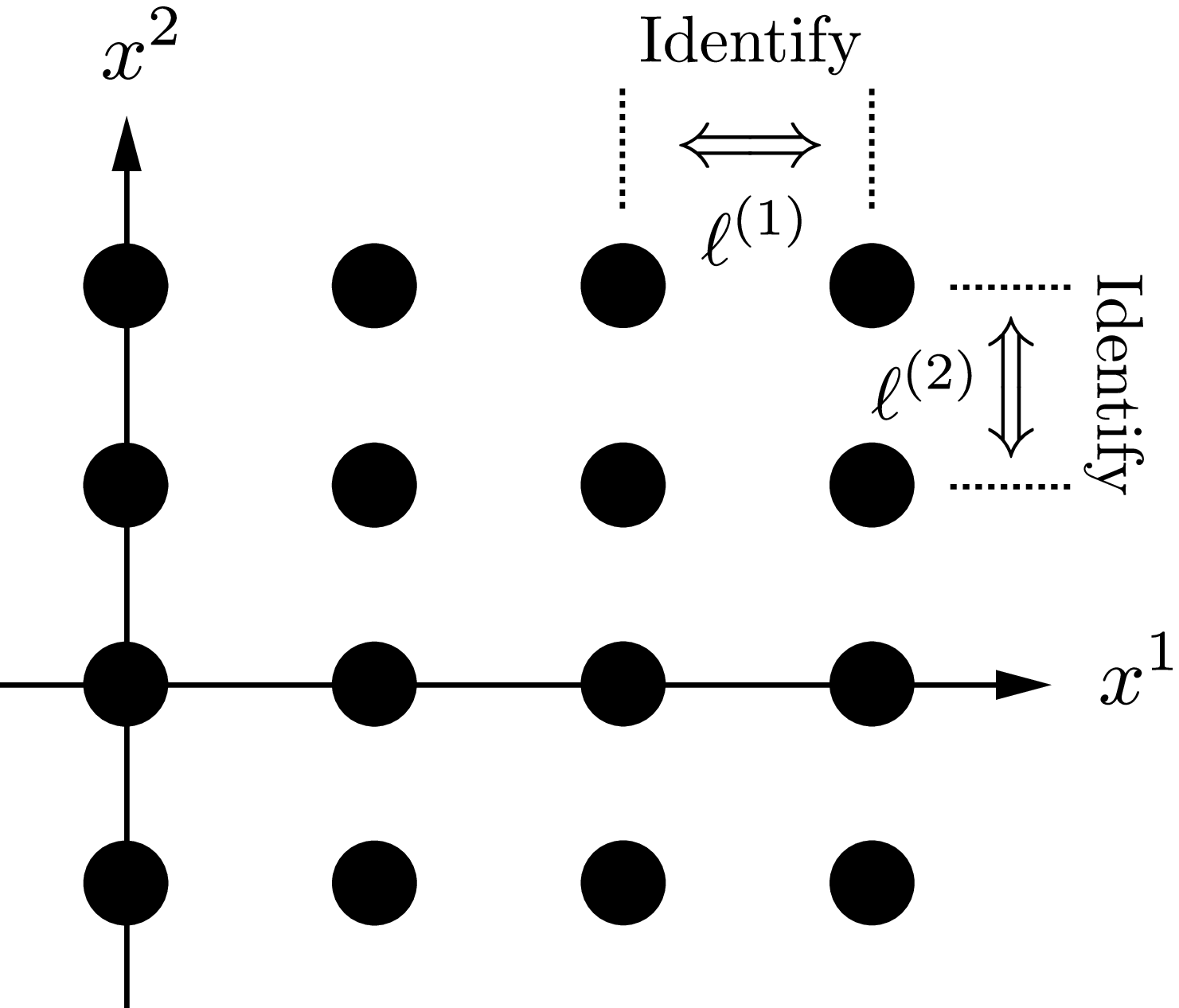}
\end{center}
\caption{
Periodic identification in the case of $T^2$ compactified extra dimensions.
}
\label{fig:t2extradim}
\end{figure}

The explicit form of the metric of Kaluza-Klein black hole is given by
\begin{eqnarray}
ds^2 &=& -H^{-2}dt^2 
+H^{2/(D-3)} \left[
\sum_{I,J=1}^{N}\delta_{IJ}dx^Idx^J
+
\sum_{A,B=N+1}^{D-1}\delta_{AB}dx^Adx^B
\right],
\label{metkkbh}
\\
H &=& 1 +  \sum_{ n_1,n_2,\cdots,n_N =-\infty}^\infty \frac{m}{
\left[
\displaystyle\sum_{I,J = 1}^{N}\delta_{IJ}(x^I - n_I \ell^{(I)})(x^J - n_J \ell^{(J)})
+
\displaystyle\sum_{A,B = N+1}^{D-1} \delta_{AB}x^A x^B
\right]^{(D-3)/2}},
\notag
\end{eqnarray}
where  $\ell^{(I)}$ denotes the size of extra dimension in the $x^I$ direction,
and the number of the compactified extra dimensions $N$ satisfies $1 \le N \le D-4$.
(see Fig.\ref{fig:t2extradim} in the case of $N=2$ for example).

\subsection{$C^1$ extension across the horizon}
We 
can place the black hole at the origin of Euclidean coordinates without loss of generality.
Let us introduce polar coordinates as
\begin{eqnarray}
x^i &=& r 
\left[\prod_{j=1}^{i-1}  \sin \theta^j \right]   \cos \theta^i
~(=: r \Theta^i )~~(1 \le i \le D-2),
\\
x^{D-1} &=& r 
\left[\prod_{j=1}^{D-2}  \sin \theta^j \right]
(=: r \Theta^{D-1} ).
\end{eqnarray}
Then, the metric (\ref{metkkbh}) becomes
\begin{eqnarray}
ds^2 &=& -H^{-2} dt^2 + H^{2/(D-3)}(dr^2 + r^2 d\Omega_{{\rm S}^{D-2}}^2),
\\
d\Omega_{{\rm S}^{D-2}}^2 &=& \sum_{i=1}^{D-2} \left[
\prod_{j=1}^{i-1} \sin^2 \theta^j
\right]
 (d\theta^{i})^2,
\\
H &=& 1 + \frac{m}{r^{D-3}} 
+  m \sum_{n_1,n_2,\cdots,n_N \neq \{0,0,\cdots,0\}} (H_{n_1,n_2,\cdots,n_N})^{-(D-3)/2},
\\
H_{n_1,n_2,\cdots,n_N} &=& 
\sum_{I,J = 1}^{N}\delta_{IJ}(r \Theta^I - n_I \ell^{(I)})(r \Theta^J - n_J \ell^{(J)})
+
\sum_{A,B = N+1}^{D-1} \delta_{AB}r^2 \Theta^A  \Theta^B.
\end{eqnarray}
In these coordinates, the metric component $g_{rr}$ diverges at the horizon $r=0$.
To 
remove this divergence, we introduce
a coordinate $u$ as
\begin{eqnarray}
dt 
&=&
du 
-
H^{(D-2)/(D-3)} dr 
+
\sum_{i = 1}^{D-2}Y_i d\theta^i,
\label{EFcoord1}
\\
Y_i &:=& -\int dr \frac{\partial H^{(D-2)/(D-3)}}{\partial \theta^i}.
\end{eqnarray}
Note that the last term in Eq.~(\ref{EFcoord1}) is needed to satisfy the integrability condition.
Near $r = 0$, we can show that
\begin{eqnarray}
Y_i &\simeq & W_i(\theta) r^2,
\label{eqyi}
\end{eqnarray}
where $W_i(\theta)$ is a finite function of the angular coordinates.
To derive Eq.~(\ref{eqyi}), we used the condition that the function $\sum (H_{n_1,n_2,\cdots,n_N})^{-(D-3)/2} $ is a function of $r^2$ and angular coordinates
\begin{eqnarray}
\sum_{n_1,n_2,\cdots,n_N \neq \{0,0,\cdots,0\}} (H_{n_1,n_2,\cdots,n_N})^{-(D-3)/2} = C_0 + C_2(\theta) r^2 + C_4(\theta) r^4 + \cdots
\end{eqnarray}
since the function $\sum (H_{n_1,n_2,\cdots,n_N})^{-(D-3)/2} $ has 
formally $r \leftrightarrow -r$ symmetry.
Then, the metric becomes 
\begin{eqnarray}
ds^2 &=&
 -H^{-2} du^2 
 -
H^{-2}\sum_{i,j}^{D-2}Y_i Y_j d\theta^i d\theta^j 
+
2  H^{-(D-4)/(D-3)} du dr
-
2 H^{-2} \sum_{i}^{D-2}Y_i  du  d\theta^i
\notag\\ & &
+
2 
H^{-(D-4)/(D-3)}  \sum_{i}^{D-2} Y_i dr d\theta^i
+
H^{2/(D-3)}
 r^2 d\Omega^2_{S^{D-2}}.
\end{eqnarray}
However, in these coordinates, all the metric components except 
for the coefficient of $d\Omega^2$ become zero
at the horizon because 
the functions $H, Y_i$ behaves as
\begin{eqnarray}
H & \sim & r^{-(D-3)},
\\
Y_i &\sim & r^{2},
\end{eqnarray}
near $r=0$.
Thus, the metric in these coordinates degenerates at the horizon $r=0$.
We can remove this coordinate singularity by 
further introducing a new radial coordinate $\rho$ as
\begin{eqnarray}
\rho = r^{D-3}.
\end{eqnarray}
Then, the metric behaves as
\begin{eqnarray}
ds^2 &=&
2(D-3)^{-1} m^{-(D-4)/(D-3)}   
  du d\rho
+
2  \sum_{i}^{D-2} F_i d\rho d\theta^i
+
m^{2/(D-3)}  d\Omega^2_{S^{D-2}} + {\cal O}(\rho),
\label{rthetacomp}
\\
F_i &=&
(D-3)^{-1}  m^{-(D-4)/(D-3)} Y_i
\notag\\&=&
(D-3)^{-1}  m^{-(D-4)/(D-3)} \left(
W_i(\theta) \rho^{2/(D-3)}  + {\cal O}(\rho^{4/(D-3)})
\right).
\end{eqnarray}
{}From Eq.~(\ref{rthetacomp}), 
we can see the function $\partial_\rho g_{\rho \theta^i}$ diverges at the horizon,
thus  this extension is only a $C^0$ extension, not being a $C^1$ extension, if $D \ge 6$.
To get a $C^1$ extension, we introduce new coordinates $\bar{u}$ and $\bar{\theta}^i$ as
\begin{eqnarray}
d \bar{\theta}^i &:= & 
d\theta^i +
\frac{1}{m^{2/(D-3)}}
{\cal G}^i d\rho + 
\frac{1}{m^{2/(D-3)}}
\sum_{j} d\theta^j \int d\rho \partial_{\theta^j}{\cal G}^i,
\label{thetabar}
\\
d\bar{u} &:= & 
(D-3)^{-1} m^{-(D-4)/(D-3)}   du 
-
\frac{1}{m^{2/(D-3)}}
\sum_{i=1}^{D-2} 
\frac{\left({\cal G}^i \right)^2}{2} \prod_{j=1}^{i-1} \sin^2 \theta^j d\rho 
\notag\\&&- 
\frac{1}{m^{2/(D-3)}}
\sum_{i,j=1}^{D-2} d\theta^j \left(\int d\rho \partial_{\theta^j}  
 \left[
\frac{\left({\cal G}^i \right)^2}{2} \prod_{j=1}^{i-1} \sin^2 \theta^j
\right]   \right),
\label{ubar}
\\
{\cal G}^i &:=&  F_i 
 \left[
\prod_{j=1}^{i-1} \sin^2 \theta^j
\right]^{-1}.
\end{eqnarray}
The dominant parts of the last terms in Eqs.~(\ref{thetabar}) and (\ref{ubar}) 
have positive powers of $\rho$ higher than unity because their integrands 
are proportional to positive fractional powers of $\rho$ near $\rho = 0$.
Then, the metric behaves as
\begin{eqnarray}
ds^2 =
 2  d\bar{u} d\rho
+ m^{2/(D-3)}(d\bar{\theta}^2 + {\sin}^2 \bar{\theta} d\Omega_{S^{D-3}}^2)
+{\cal O }(\rho^1).
\end{eqnarray}
As we find that
the derivatives of all the metric components w.r.t. $\rho$ take finite values at the horizon,
this is a $C^1$ extension across the horizon.

\subsection{Divergence of Riemann tensor in the coordinates of the $C^1$ extension}
We consider following two 1-forms
\begin{eqnarray}
e^{(0)} &=& -dt - \sqrt{-1 + H^2} H^{1/(D-3)} dr,
\\
e^{(1)} &=&  d\theta^1.
\end{eqnarray}
If we express these 1-forms by using $(d\bar{u}, d\rho, d\bar{\theta}^i)$, they become
\begin{eqnarray}
e^{(0)} &=& -(D-3)m^{(D-4)/(D-3)}d\bar{u} 
+
\frac{1}{2}m^{-(D-4)/(D-3)}(D-3)^{-1}d\rho
+
{\cal O}(\rho^{2/(D-3)}),
\\
e^{(1)} &=& d\bar{\theta}^1 + {\cal O}(\rho^{2/(D-3)}).
\end{eqnarray}
Thus, if a component of a tensor projected to $e^{(0)}$ and $e^{(1)}$ diverges,
we can say that some component of such a tensor
also diverges at $\rho = 0$ in the coordinate $(d\bar{u}, d\rho, d\bar{\theta}^i)$. 
For this reason, 
it is sufficient to confirm the divergence of the Riemann tensor projected to $e^{(0)}$ and $e^{(1)}$.
After some calculations, for $D >5$, we obtain
\begin{eqnarray}
e^{(0)}_{\mu}
e^{(1)}_{\nu}
e^{(0)}_{\rho}
e^{(1)}_{\sigma}
R^{\mu \nu \rho \sigma} &\propto & \frac{1}{\rho^{(D-5)/(D-3)}} 
\to \infty.
\end{eqnarray}
Thus, we conclude that there is no $C^2$ extension across the horizon from the theorem.~1.
By contrast, for $D = 5$, the horizon becomes analytic as shown in Ref.~\cite{Candlish:2007fh}.

\section{summary and discussion}

In this paper, we have proposed a simple method to prove the non-smoothness of the horizon
and applied it to several black hole spacetimes
for which the scalar invariants constructed from the Riemann tensor,
{\it e.g.}, the Ricci scalar and the Kretschmann invariant, 
take finite values at the horizon.
In Secs.~III and IV, we have shown that our method works well for 
a toy model and the multi black holes,
reproducing the results in Refs.~\cite{Welch:1995dh, Candlish:2007fh}.
We have shown that the Kaluza-Klein black holes constructed by Myers 
have a curvature singularity at the horizon if $D \ge 6$ in Sec.~V.

Though one may think that the existence of the curvature singularities 
immediately means the breakdown of the classical theory,
in fact, it depends on the strength of the curvature singularities.
Using our method, one can also discuss the strength of the curvature singularities.
In the case of Kaluza-Klein black holes, the Riemann tensor diverges 
as $\rho^{-(D-5)/(D-3)}$ where $\rho$ is approximately the proper length from the horizon.
In this case, since the singularity is relatively mild,
{\it i.e.}, the second integral of the Riemann tensor is finite,
the tidal force on a finite-sized body is not divergent across the horizon.

One of the advantage of our method is that
it applies to less-symmetric spacetimes, but
it has a merit even in the case of symmetric spacetimes.
Even if it is shown that 
there exists no smooth extension across the horizon
which maintains the spacetime symmetry,
there is a possibility that we may find a smoother extension
by considering an extension which breaks the
symmetry of the spacetime, like in the case of AdS Poincar\'e horizon
where we need to introduce a new coordinate system across the horizon
which does not have the same Killing coordinate of the Poincar\'e chart.
Our method can be used to prove that such a possibility is excluded.

We comment on the restriction that we have only focused on
$C^2$ extensions across the ``{\it same}'' boundary in this paper.
First, we should emphasize that our method is a natural extension of the 
previous works~\cite{Candlish:2007fh, Candlish:2009vy, Gowdigere:2014aca}.
Since the discussion in the previous works is based on the explicit construction of 
extensions across the horizon by using the same coordinate system for the outside and on the horizon,
what were discussed are in fact extensions across the ``{\it same}'' boundary.
Secondly, from the definition of the extensions across the ``{\it same}'' boundary,
if there exists any other $C^2$ extension across the boundary,
some geodesic cannot reach the boundary\footnote{
For example, there is a possibility that a geodesic oscillates infinitely many times
near the boundary and does not have a limiting point on the boundary.}
and is inextensible while the affine parameter is finite there.
This implies that there exists a singularity in the same sence as used in the singularity theorem~\cite{Hawking:1973uf}.
One may think that 
the divergence of the Riemann tensor only in a $C^1$ extension 
does not have a covariant meaning.
However, even in that case, we can say that there exists
some singularity at least in any extension from our method.

As far as we know, there is no discussion on the connection between the existence of 
p.p.~curvature singularity and the no existence of the $C^2$ extension 
for the case of Lorentzian signature
in literature. 
In lemma.~1, we have discussed it when 
there exists a p.p.~curvature singularity along a time like geodesic.

Finally, we should note that we need a $C^1$ extension across the boundary 
to give a criteria for non-existence of $C^{N+2}$ extension in our theorem.
There are possible cases that spacetimes do not admit $C^1$ extension across 
the boundary.
It is also important to study the existence and construction of $C^1$
extension for general spacetime. We leave this problem for future work.

While this paper was being prepared for submission, 
an interesting paper~\cite{Gowdigere:2014cqa} appeared, in which the 
smoothness of horizons in the most generic multi center 
black hole and membrane solutions were discussed.
\\\\

\section*{Acknowledgements}
The authors thank 
P.~Chrusciel, T.~Houri, H.~Kodama, K.~-i.~Nakao, H.~Reall, R.~Saito, T.~Shiromizu, 
K.~Tanabe, N.~Tanahashi, T.~Tatsuoka, B.~Way, Y.~Yasui and C.~-M.~Yoo
for very helpful comments and suggestions. 
MK is supported by a grant for research abroad from JSPS.
HI is supported by the Grant-in-Aid for Scientific Research No. 19540305 and 24540282.
TT is supported by the Grant-in-Aid for Scientific Research No. 26287044, 24103006 and 24103001.

\appendix

\section{behavior of $dy^\mu/d\lambda$ near ${\cal H^{\prime \prime}}$}
\label{appendixA}

In this section, we study the behavior of the tangent vector $dy^\mu/d\lambda$ in Eq.~(\ref{geoeqc2extension})
in the limit to ${\cal H^{\prime \prime}}$, {\it i.e.}, $\lambda \to \lambda_f$.
Firstly, to gain an intuitive understanding, 
we show that 
 $dy^\mu/d\lambda$ cannot diverge as a power low of the affine parameter.
Later, we treat the general case.

\subsection{the case of power low divergence}
We assume the leading behavior of 
the most divergent component of the tangent vector $dy^\mu/d\lambda$ near $p^{\prime \prime}$ as
\begin{eqnarray}
\frac{dy^\mu}{d\lambda} \sim \frac{c}{|\lambda - \lambda_f|^{m}},
\label{appendix_D_1_01}
\end{eqnarray}
where $c~(\neq 0)$ and $m~(> 0)$ are constants.
Since the coordinates  $\{y^\mu \}$ cover the point $p^{\prime \prime}$, 
the values of $y^\mu$ at $p^{\prime \prime}$ should be finite.
Then, the power $m$ in Eq.~(\ref{appendix_D_1_01}) should be less than unity
\begin{eqnarray}
m < 1.
\end{eqnarray}
Substituting this to Eq.~(\ref{geoeqc2extension}), we find that
the leading behaviors of the left-hand side~(LHS) and right-hand side~(RHS) become
\begin{eqnarray}
{\rm LHS} &\sim& \frac{c}{|\lambda - \lambda_f|^{m+1}},
\\
|{\rm RHS}| & \le & |\Gamma^\mu{}_{\alpha \beta}| \frac{c^2}{|\lambda - \lambda_f|^{2m}}.
\end{eqnarray}
Since $m <1$ and the Christoffel symbols take finite values, 
the LHS cannot be balanced with RHS in Eq.~(\ref{geoeqc2extension}). This contradiction shows that our assumption~(\ref{appendix_D_1_01})
cannot be true.

\subsection{the general case}
{}

Without loss of generality, we can assume
that the tangent of the geodesic is future-directed.\footnote{
If the tangent of the geodesic is past-directed,
we only have to interchange the past and the future in the following discussion.}
From Appendix.~\ref{appendixD} we can assume that all
vectors normal to the constant surfaces 
of the coordinate functions $y^\mu$ are timelike and
future directed, at least, in the vicinity of a point $p^{\prime \prime}$ on ${\cal H}^{\prime \prime}$.
In this case, the tangent of a time like geodesic ending at $p^{\prime \prime}$
satisfies
\begin{eqnarray}
\frac{dy^\mu}{d\lambda} > 0,
\end{eqnarray}
as long as a sufficiently short geodesic is concerned.
We introduce a non-affine parameter $\zeta$ 
for the geodesic $y^\mu(\lambda)$
increasing toward the future satisfying
\begin{eqnarray}
\delta_{\mu \nu}\frac{dy^\mu}{d\zeta}\frac{dy^\nu}{d\zeta} = 1,
\label{zetaparameter}
\end{eqnarray}
then, we also have
\begin{eqnarray}
\frac{dy^\mu}{d\zeta} > 0.
\end{eqnarray}
{}From these equations we can show that the value of the parameter $\zeta$ at 
$p^{\prime \prime}$ on ${\cal H^{\prime \prime}}$
has a definite value as
\begin{eqnarray}
\zeta(p^{\prime \prime})
-
\zeta|_{\lambda = \lambda_i}
&=& 
\int_{\lambda_i}^{\lambda_f} d\lambda
\sqrt{\delta_{\mu \nu}\left|\frac{dy^\mu}{d\lambda}\right|\left|\frac{dy^\nu}{d\lambda}\right|}
\notag\\&< & 
\int_{\lambda_i}^{\lambda_f} d\lambda
\sum_\mu \left|\frac{dy^\mu}{d\lambda}\right|
\notag\\&= & 
\int_{\lambda_i}^{\lambda_f}  d\lambda
\sum_\mu \frac{dy^\mu}{d\lambda}
\notag\\&= & 
\sum_\mu 
(y^\mu|_{p^{\prime \prime}} - y^\mu|_{\lambda = \lambda_i})
\notag\\ &<& \infty.
\end{eqnarray}
For any finite value of $\zeta|_{\lambda = \lambda_i}$, 
$\zeta(p^{\prime \prime})$ is finite.
The geodesic equation in term of the parameter $\zeta$ becomes
\begin{eqnarray}
\frac{d^2y^\mu}{d\zeta^2} 
=
-
 \Gamma^\mu{}_{\alpha \beta}
\frac{dy^\alpha}{d\zeta} 
  \frac{dy^\beta}{d\zeta} 
+
\delta_{\rho \nu}
\Gamma^\rho{}_{\alpha \beta}
\frac{dy^\alpha}{d\zeta}
\frac{dy^\beta}{d\zeta}
\frac{dy^\nu}{d\zeta} 
\frac{dy^\mu}{d\zeta}.
\end{eqnarray}
Since $dy^\mu/d\zeta$ does not diverge at $p^{\prime \prime}$ from 
the definition of the parameter $\zeta$ in Eq.~(\ref{zetaparameter}),
 $d^2y^\mu/d\zeta^2$ also does not diverge at $p^{\prime \prime}$.
For this reason, all components of $dy^\mu/d\zeta$ have definite values $w^\mu$ at 
$p^{\prime \prime}$.\footnote{
Since the integrand is bounded above and below,
$dy^\mu/d\zeta = \int d\zeta d^2y^\mu/d\zeta^2$
takes definite value.
{}From the Eq.~(\ref{zetaparameter}), we can say that $w^\mu$ cannot be zero vector.
}

On the other hand, when we consider the solution $\bar{y}^\mu(\bar{\lambda})$ of the equation 
\begin{eqnarray}
\frac{d}{d\bar{\lambda}} \frac{d\bar{y}^\mu}{d\bar{\lambda}} + \Gamma^\mu{}_{\alpha \beta}
\frac{d\bar{y}^\alpha}{d\bar{\lambda}}
\frac{d\bar{y}^\beta}{d\bar{\lambda}}
=0,
\end{eqnarray}
with the initial conditions
\begin{eqnarray}
\frac{d\bar{y}^\mu}{d\bar{\lambda}}
=
w^\mu,
\end{eqnarray}
at the point $p^{\prime \prime}$,
in a similar manner, with the parameter $\bar{\zeta}$ satisfying
\begin{eqnarray}
\delta_{\mu \nu}\frac{d\bar{y}^\mu}{d\bar{\zeta}}\frac{d\bar{y}^\nu}{d\bar{\zeta}} = 1,
\label{barzetaparameter}
\end{eqnarray}
we can show
$d\bar{y}^\mu / d\bar{\zeta} $ take definite values $\bar{w}^\mu$ at the point $p^{\prime \prime}$.
{}From the relation
\begin{eqnarray}
\frac{d\bar{y}^\mu}{d\bar{\lambda}} = 
\frac{d\bar{\zeta}}{d\bar{\lambda}}
\frac{d\bar{y}^\mu}{d\bar{\zeta}},
\end{eqnarray}
 we have a relation 
\begin{eqnarray}
 w^{\mu} = 
\frac{d\bar{\zeta}}{d\bar{\lambda}}\bigg|_{p^{\prime \prime}}
\bar{w}^\mu.
\end{eqnarray}
From the definition of $w^{\mu}$ and $\bar{w}^\mu$,
we have relations 
\begin{eqnarray}
\delta_{\mu \nu}w^{\mu}w^{\nu} &=& 1,
\\
\delta_{\mu \nu}\bar{w}^\mu \bar{w}^\nu &=& 1.
\end{eqnarray}
Thus, we conclude $d\bar{\zeta}/d\bar{\lambda}|_{p^{\prime \prime}} = 1$,
and it is clear that 
the orbit $y^\mu = \bar{y}^\mu(\bar{\lambda}) $ is the same as $y^\mu(\lambda)$ or $y^\mu(\zeta)$.

Two affine parameters of
the same orbit must be related by an affine transformation
\begin{eqnarray}
\bar{\lambda} = \alpha \lambda + \beta,
\end{eqnarray}
and the tangent vectors are related as
\begin{eqnarray}
\frac{dy^\mu}{d\lambda}
=
\alpha 
\frac{dy^\mu}{d\bar{\lambda}} 
\stackrel{p^{\prime \prime}}{\to} \alpha w^\mu.
\end{eqnarray}
If a component of $dy^\mu/d\lambda$ diverges at the point $p^{\prime \prime}$,
the only possibility is $\alpha \to \infty$.
However, in that case, 
it diverges everywhere on the curve.
This contradicts the assumption that $dy^\mu/d\lambda$  does not diverge in
$\mu^{\prime \prime}({\cal M})$.

\section{parallel transport of vector along a geodesic in $C^1$ spacetime}
\label{appendixB}
In this section, let the spacetime $(M^\prime, g_{\mu \nu}^\prime)$
be a $C^2$ manifold with a $C^1$ metric tensor.
We focus on a chart described by local coordinates $\{x^\mu \}$ on ${\cal M^\prime}$, and 
we assume that there exists a geodesic $\gamma^\prime(\lambda)$ on this chart.
Denoting the geodesic $\gamma^\prime(\lambda)$ by $x^\mu(\lambda)$,
we find that
the function $x^\mu(\lambda)$ is twice differentiable owing to the geodesic equation.
Let us consider 
a vector $e^\mu_0$ at a point $\gamma^\prime(\lambda_0)$ on the geodesic and 
its parallel transport $e^\mu$ along the geodesic.
To find the parallelly transported vector field $e^\mu$ on $\gamma^\prime(\lambda)$, 
we should solve the differential equation
\begin{eqnarray}
\frac{d e^\nu}{d\lambda}
+
\Gamma^\nu{}_{\rho \sigma}
\frac{dx^\rho}{d\lambda}
e^\sigma
= 0.
\end{eqnarray}
for $e^\mu$ with the initial value $e^\mu = e^\mu_0$ at $\lambda = \lambda_0$.
Since all components of the Christoffel symbol, which take finite values on ${\cal M^\prime}$, 
can be considered as functions of $\lambda$ along the curve $x^\mu(\lambda)$, 
this equation is a system of linear ordinary differential equations with finite coefficients.
Thus, the solution for a given initial data is unique and 
finite as long as $\lambda$ is finite.

\section{geodesics starting with ${\cal H^\prime}$}
\label{appendixC}

In this section we discuss the existence of a solution of the geodesic equation
emanating from a point on ${\cal H^\prime}$ in $({\cal M^\prime}, g_{\mu \nu}^\prime)$,
where 
${\cal M^\prime}$ is a $C^2$ manifold and the metric $g_{\mu \nu}^\prime$ is $C^1$.
The geodesic equation is given by
\begin{eqnarray}
\frac{dx^\mu}{d\lambda} &=& v^\mu,
\label{appendix_B_01}
\\
\frac{d v^\mu}{d\lambda} &=& 
- \Gamma^\mu{}_{\alpha \beta}
v^\alpha
v^\beta.
\label{appendix_B_02}
\end{eqnarray}
Since the Christoffel symbol is continuous, we can say that 
there exists at least one solution for any initial condition from the Peano existence theorem.

We can find a solution iteratively for an initial values $x^\mu = x^\mu_{\rm ini}, v^\mu = v^\mu_{\rm ini}$
unless all the components of $v^\mu_{\rm ini}$ vanish.
Firstly, we solve the Eqs.~(\ref{appendix_B_01}) and (\ref{appendix_B_02}) approximately as
\begin{eqnarray}
x^\mu(\lambda) &\simeq & x_{\rm 1st}^\mu(\lambda)
\notag\\
&:=& x^\mu_{\rm ini} +  v^\mu_{\rm ini} \lambda,
\label{appendix_B_03}
\\
v^\mu(\lambda) 
& \simeq & v^\mu_{\rm 1st}(\lambda)
\notag\\
&:= &
 v^\mu_{\rm ini} 
- \Gamma^\mu{}_{\alpha \beta}(x^\nu_{\rm ini} )
v^\alpha_{\rm ini}
v^\beta_{\rm ini} \lambda.
\label{appendix_B_04}
\end{eqnarray}
Substituting these to the right hand side of the Eqs.~(\ref{appendix_B_01}) and (\ref{appendix_B_02}) again,
we obtain the next order approximation as
\begin{eqnarray}
x^\mu(\lambda) &\simeq & x_{\rm 1st}^\mu(\lambda)
+
 x_{\rm 2nd}^\mu(\lambda),
\notag\\
&:=& x^\mu_{\rm ini} +  v^\mu_{\rm ini} \lambda
-\frac{1}{2} \Gamma^\mu{}_{\alpha \beta}(x^\nu_{\rm ini} )
v^\mu_{\rm ini}
v^\mu_{\rm ini} \lambda^2,
\label{appendix_B_05}
\\
v^\mu(\lambda) 
& \simeq & v^\mu_{\rm 1st}(\lambda)
+
v^\mu_{\rm 2nd}(\lambda)
\notag\\
&:= &
 v^\mu_{\rm ini} 
- \Gamma^\mu{}_{\alpha \beta}(x^\nu_{\rm ini} )
v^\mu_{\rm ini}
v^\mu_{\rm ini} \lambda
 +
\int_0^\lambda d\lambda^\prime
\bigg[
2
\Gamma^\mu{}_{\alpha \beta}(x^\nu_{\rm 1st}(\lambda^\prime))
\Gamma^\beta{}_{\rho \sigma}(x^\nu_{\rm ini} )
v^\alpha_{\rm ini}
v^\rho_{\rm ini}
v^\sigma_{\rm ini} \lambda^\prime
\notag
\\&& -
\Gamma^\mu{}_{\alpha \beta}(x^\nu_{\rm 1st}(\lambda^\prime))
\Gamma^\alpha{}_{\rho \sigma}(x^\nu_{\rm ini} )
\Gamma^\beta{}_{\kappa \mu}(x^\nu_{\rm ini} )
v^\rho_{\rm ini}
v^\sigma_{\rm ini}
v^\kappa_{\rm ini}
v^\mu_{\rm ini} \lambda^{\prime 2}
\bigg].
\label{appendix_B_06}
\end{eqnarray}
Repeating this process, we can obtain a solution of geodesic equation locally.

\section{coordinate system in which all coordinates are time coordinates}
\label{appendixD}

If both the manifold and metric are $C^2$, we can introduce Riemann normal 
coordinates around any point $p$. 
In these coordinates
the metric becomes like
\begin{eqnarray}
ds^2 = -dt^2 + \sum_{i,j}\delta_{ij}dx^idx^j 
+
\frac{1}{3} R_{\mu \alpha \beta \nu} x^\alpha x^\beta dx^\mu dx^\nu
+
{\cal O}(x^3),
\end{eqnarray}
where we choose the point $p$ as the origin of coordinates.
If we introduce a new coordinate system $\{y^\mu \}$ as
\begin{eqnarray}
y^0 &=& t,
\\
y^i &=& t + \epsilon x^i,
\end{eqnarray}
where $\epsilon$ is a constant,
the norm of the normal vector of this coordinate $y^i$ become
\begin{eqnarray}
|dy^i|^2 &=& |dt + \epsilon dx^i |^2
\notag\\&=&
-1 + \epsilon^2 + \epsilon 
\left[-\frac{1}{3}R_{0 \alpha \beta i}x^\alpha x^\beta + {\cal O}(x^3)
\right].
\end{eqnarray}
Then, if we choose $\epsilon \ll 1 $, the norm of $dy^i$ becomes negative near the point $p$.
Restricting the region of the coordinate system 
to the neighborhood in which 
all the norms of $dy^i$ take negative values,
we can obtain 
a local coordinate system around any point $p$ in which 
all normal vectors of the coordinate functions $y^\mu$ 
are timelike
and future directed.


\begin{thebibliography}{99}

\bibitem{Banks:1999gd} 
  T.~Banks and W.~Fischler,
  hep-th/9906038.

\bibitem{Dimopoulos:2001hw} 
  S.~Dimopoulos and G.~L.~Landsberg,
  Phys.\ Rev.\ Lett.\  {\bf 87}, 161602 (2001)
  [hep-ph/0106295].

\bibitem{Giddings:2001bu} 
  S.~B.~Giddings and S.~D.~Thomas,
  Phys.\ Rev.\ D {\bf 65}, 056010 (2002)
  [hep-ph/0106219].

\bibitem{Ida:2002ez} 
  D.~Ida, K.~-y.~Oda and S.~C.~Park,
  Phys.\ Rev.\ D {\bf 67}, 064025 (2003)
  [Erratum-ibid.\ D {\bf 69}, 049901 (2004)]
  [hep-th/0212108].

\bibitem{Ida:2005ax} 
  D.~Ida, K.~-y.~Oda and S.~C.~Park,
  Phys.\ Rev.\ D {\bf 71}, 124039 (2005)
  [hep-th/0503052].

\bibitem{Ida:2006tf} 
  D.~Ida, K.~-y.~Oda and S.~C.~Park,
  Phys.\ Rev.\ D {\bf 73}, 124022 (2006)
  [hep-th/0602188].



\bibitem{Argyres:1998qn} 
  P.~C.~Argyres, S.~Dimopoulos and J.~March-Russell,
  Phys.\ Lett.\ B {\bf 441}, 96 (1998)
  [hep-th/9808138].

\bibitem{Feng:2001ib} 
  J.~L.~Feng and A.~D.~Shapere,
  Phys.\ Rev.\ Lett.\  {\bf 88}, 021303 (2002)
  [hep-ph/0109106].

\bibitem{Anchordoqui:2001cg} 
  L.~A.~Anchordoqui, J.~L.~Feng, H.~Goldberg and A.~D.~Shapere,
  Phys.\ Rev.\ D {\bf 65}, 124027 (2002)
  [hep-ph/0112247].


\bibitem{Dobiasch:1981vh} 
  P.~Dobiasch and D.~Maison,
  Gen.\ Rel.\ Grav.\  {\bf 14}, 231 (1982).

\bibitem{Gibbons:1985ac} 
  G.~W.~Gibbons and D.~L.~Wiltshire,
  Annals Phys.\  {\bf 167}, 201 (1986)
  [Erratum-ibid.\  {\bf 176}, 393 (1987)].

\bibitem{Gauntlett:2002nw} 
  J.~P.~Gauntlett, J.~B.~Gutowski, C.~M.~Hull, S.~Pakis and H.~S.~Reall,
  Class.\ Quant.\ Grav.\  {\bf 20}, 4587 (2003)
  [hep-th/0209114].

\bibitem{Gaiotto:2005gf} 
  D.~Gaiotto, A.~Strominger and X.~Yin,
  JHEP {\bf 0602}, 024 (2006)
  [hep-th/0503217].

\bibitem{Ishihara:2005dp} 
  H.~Ishihara and K.~Matsuno,
  Prog.\ Theor.\ Phys.\  {\bf 116}, 417 (2006)
  [hep-th/0510094].

\bibitem{Wang:2006nw} 
  T.~Wang,
  Nucl.\ Phys.\ B {\bf 756}, 86 (2006)
  [hep-th/0605048].

\bibitem{Yazadjiev:2006iv} 
  S.~S.~Yazadjiev,
  Phys.\ Rev.\ D {\bf 74}, 024022 (2006)
  [hep-th/0605271].

\bibitem{Nakagawa:2008rm} 
  T.~Nakagawa, H.~Ishihara, K.~Matsuno and S.~Tomizawa,
  Phys.\ Rev.\ D {\bf 77}, 044040 (2008)
  [arXiv:0801.0164 [hep-th]].

\bibitem{Tomizawa:2008hw} 
  S.~Tomizawa, H.~Ishihara, K.~Matsuno and T.~Nakagawa,
  Prog.\ Theor.\ Phys.\  {\bf 121}, 823 (2009)
  [arXiv:0803.3873 [hep-th]].

\bibitem{Matsuno:2008fn} 
  K.~Matsuno, H.~Ishihara, T.~Nakagawa and S.~Tomizawa,
  Phys.\ Rev.\ D {\bf 78}, 064016 (2008)
  [arXiv:0806.3316 [hep-th]].

\bibitem{Tomizawa:2008rh} 
  S.~Tomizawa and A.~Ishibashi,
  Class.\ Quant.\ Grav.\  {\bf 25}, 245007 (2008)
  [arXiv:0807.1564 [hep-th]].

\bibitem{Stelea:2008tt} 
  C.~Stelea, K.~Schleich and D.~Witt,
  Phys.\ Rev.\ D {\bf 78}, 124006 (2008)
  [arXiv:0807.4338 [hep-th]].

\bibitem{Tomizawa:2008qr} 
  S.~Tomizawa, Y.~Yasui and Y.~Morisawa,
  Class.\ Quant.\ Grav.\  {\bf 26}, 145006 (2009)
  [arXiv:0809.2001 [hep-th]].

\bibitem{Gal'tsov:2008sh} 
  D.~V.~Gal'tsov and N.~G.~Scherbluk,
  Phys.\ Rev.\ D {\bf 79}, 064020 (2009)
  [arXiv:0812.2336 [hep-th]].

\bibitem{Bena:2009ev} 
  I.~Bena, G.~Dall'Agata, S.~Giusto, C.~Ruef and N.~P.~Warner,
  JHEP {\bf 0906}, 015 (2009)
  [arXiv:0902.4526 [hep-th]].

\bibitem{Tomizawa:2010xq} 
  S.~Tomizawa,
  arXiv:1009.3568 [hep-th].

\bibitem{Mizoguchi:2011zj} 
  S.~'y.~Mizoguchi and S.~Tomizawa,
  Phys.\ Rev.\ D {\bf 84}, 104009 (2011)
  [arXiv:1106.3165 [hep-th]].

\bibitem{Chen:2010ih} 
  Y.~Chen and E.~Teo,
  Nucl.\ Phys.\ B {\bf 850}, 253 (2011)
  [arXiv:1011.6464 [hep-th]].

\bibitem{Stelea:2011fj} 
  C.~Stelea, K.~Schleich and D.~Witt,
  arXiv:1108.5145 [gr-qc].

\bibitem{Nedkova:2011hx} 
  P.~G.~Nedkova and S.~S.~Yazadjiev,
  Phys.\ Rev.\ D {\bf 84}, 124040 (2011)
  [arXiv:1109.2838 [hep-th]].

\bibitem{Tatsuoka:2011tx} 
  T.~Tatsuoka, H.~Ishihara, M.~Kimura and K.~Matsuno,
  Phys.\ Rev.\ D {\bf 85}, 044006 (2012)
  [arXiv:1110.6731 [hep-th]].

\bibitem{Nedkova:2011aa} 
  P.~G.~Nedkova and S.~S.~Yazadjiev,
  Phys.\ Rev.\ D {\bf 85}, 064021 (2012)
  [arXiv:1112.3326 [hep-th]].

\bibitem{Mizoguchi:2012vg} 
  S.~'y.~Mizoguchi and S.~Tomizawa,
  Phys.\ Rev.\ D {\bf 86}, 024022 (2012)
  [arXiv:1201.3063 [hep-th]].



\bibitem{Myers:1986rx}
  R.~C.~Myers,
  Phys.\ Rev.\  D {\bf 35}, 455 (1987).



\bibitem{Majumdar:1947eu}
  S.~D.~Majumdar,
  Phys.\ Rev.\  {\bf 72}, 390 (1947).

\bibitem{Papaetrou:1947ib}
  A.~Papaetrou,
  Proc.\ Roy.\ Irish Acad.\ (Sect.\ A) A {\bf 51} (1947) 191.

\bibitem{Hartle:1972ya}
  J.~B.~Hartle and S.~W.~Hawking,
  Commun.\ Math.\ Phys.\  {\bf 26}, 87 (1972).

\bibitem{Candlish:2007fh}
  G.~N.~Candlish and H.~S.~Reall,
  Class.\ Quant.\ Grav.\  {\bf 24}, 6025 (2007)
  [arXiv:0707.4420 [gr-qc]].

\bibitem{Gibbons:1994vm} 
  G.~W.~Gibbons, G.~T.~Horowitz and P.~K.~Townsend,
  Class.\ Quant.\ Grav.\  {\bf 12}, 297 (1995)
  [hep-th/9410073].

\bibitem{Welch:1995dh}
  D.~L.~Welch,
  Phys.\ Rev.\  D {\bf 52}, 985 (1995)
  [arXiv:hep-th/9502146].

\bibitem{Candlish:2009vy}
  G.~N.~Candlish,
  arXiv:0904.3885 [hep-th].

\bibitem{Kimura:2008cq}
  M.~Kimura,
  Phys.\ Rev.\  D {\bf 78}, 047504 (2008)
  [arXiv:0805.1125 [gr-qc]].

\bibitem{Ishihara:2006iv}
  H.~Ishihara, M.~Kimura, K.~Matsuno and S.~Tomizawa,
  Class.\ Quant.\ Grav.\  {\bf 23}, 6919 (2006)
  [arXiv:hep-th/0605030].

\bibitem{Ishihara:2006pb}
  H.~Ishihara, M.~Kimura, K.~Matsuno and S.~Tomizawa,
  Phys.\ Rev.\  D {\bf 74}, 047501 (2006)
  [arXiv:hep-th/0607035].

\bibitem{Gowdigere:2014aca} 
  C.~N.~Gowdigere, A.~Kumar, H.~Raj and Y.~K.~Srivastava,
  arXiv:1401.5189 [hep-th].

\bibitem{Chrusciel:1992tj} 
  P.~T.~Chrusciel and D.~B.~Singleton,
  Commun.\ Math.\ Phys.\  {\bf 147}, 137 (1992).



\bibitem{Hawking:1973uf} 
  S.~W.~Hawking and G.~F.~R.~Ellis,
  ``The Large scale structure of space-time,''
  Cambridge University Press, Cambridge, 1973


\bibitem{Gowdigere:2014cqa} 
  C.~N.~Gowdigere,
  arXiv:1407.5338 [hep-th].

\end{thebibliography}
\end{document}